\documentclass[12pt,showpacs,showkeys,amsmath,amssymb]{revtex4}
\usepackage{amsmath,amsfonts,amsthm,amscd,amssymb,latexsym}
\usepackage{bm}
\usepackage{dcolumn}
\usepackage{graphicx}
\usepackage{epstopdf}
\usepackage{color}
\usepackage{epsf}
\usepackage{epsfig}
\usepackage{graphicx, epic, eepic, color}
\usepackage{xcolor}
\newcommand{\beq}{\begin{equation}}
	\newcommand{\eeq}{\end{equation}}

\def\@{\partial_}
\def\be{\begin{equation}}
	\def\ee{\end{equation}}

\def\negenspace{\kern-1.1em}

\def\sqr#1#2{{\vcenter{\hrule height.#2pt\hbox{\vrule width.#2pt
				height#1pt \kern#1pt \vrule width.#2pt}\hrule height.#2pt}}}

\begin{document}
	
\title{Local Limit of Nonlocal Gravity: Cosmological Perturbations}

\author{Javad \surname{Tabatabaei}$^{1}$}
\email{smj_tabatabaei@physics.sharif.edu}
\author{Abdolali \surname{Banihashemi}$^{1}$}
\email{abdolali.banihashemi@sharif.edu}
\author{Shant \surname{Baghram}$^{1}$}
\email{baghram@sharif.edu}
\author{Bahram \surname{Mashhoon}$^{1,2,3}$} 
\email{mashhoonb@missouri.edu}

\affiliation{
$^1$Department of Physics, Sharif University of Technology, Tehran 11155-9161, Iran\\
$^2$School of Astronomy,
Institute for Research in Fundamental Sciences (IPM),
Tehran 19395-5531, Iran\\
$^3$Department of Physics and Astronomy,
University of Missouri, Columbia,
Missouri 65211, USA
}

\date{\today}

\begin{abstract}
We explore the cosmological implications of the local limit of nonlocal gravity, which is a classical generalization of Einstein's theory of gravitation within the framework of teleparallelism. An appropriate solution of this theory is the modified Cartesian flat cosmological model. The main purpose of this paper is to  study linear  perturbations about the orthonormal tetrad frame field adapted to the standard comoving observers in this model. The observational viability of the perturbed model is examined using all available data regarding the cosmic microwave background. The implications of the linearly perturbed modified  Cartesian flat model are examined and it is shown that the model is capable of alleviating the $H_0$ tension.
\end{abstract}


\maketitle

% @@@@@@@@@@@@@@@@@@@
% @@@@@@@@@@@@@@@@@@@
% @@@@@@@@@@@@@@@@@@@
% @@@@@@@@@@@@@@@@@@@
\section{Introduction}
\label{sec1}

Cosmological observations can, in general, be well described by the standard $\Lambda$CDM model, where the cosmological constant $\Lambda$ is assumed to be the cause of the accelerated expansion of the universe (``dark energy") and the cold dark matter (CDM) is supposed to be the main component of the cosmic web. The universe, in terms of its energy content, consists of about $70\%$ dark energy, about $25\%$ dark matter and about $5\%$ visible matter. On the other hand, dark aspects of the universe provide the impetus to modify Einstein's theory of gravitation~\cite{Einstein}. 

There have been many attempts at modifications of  Einstein's general relativity (GR). For a review of the latest developments in modified gravity theories see~\cite{Sotiriou:2008rp,DeFelice:2010aj,MGC, Capozziello:2021krv}. We are interested in a classical nonlocal generalization of GR; that is, nonlocal gravity (NLG) theory, which modifies one of the fundamental pillars of relativity theory, namely, the \emph{locality hypothesis}. The gravitational nonlocality is implemented within the framework of the teleparallel equivalent of GR. This framework presents many similarities between gravitation and electromagnetism (EM). The constitutive relations in EM, which incorporate the effect of medium on the EM fields, are in general nonlocal. In a similar way, nonlocality could be introduced into the teleparallel equivalent of GR.  In electrodynamics, nonlocal constitutive relations could be approximated by simple local relations. We follow this approach in the present paper and consider the local limit of nonlocal gravity and investigate its cosmological implications.  The vast amount of cosmological data from cosmic microwave background (CMB) \cite{Planck:2018vyg} to large-scale structure (LSS) survey data \cite{SDSS:2005xqv} are essential to studying the nature of gravitation beyond the Milky Way \cite{Weinberg:2013agg} and also to constrain modified gravity models \cite{Song:2006ej,Zhang:2007nk,Baghram:2009fr,Samushia:2012iq}. The modified Friedmann-Lema\^itre-Robertson-Walker (FLRW) models were studied in the local limit of NLG in~\cite{Tabatabaei:2022tbq} and the modified Cartesian flat model was singled out as being the most plausible. The purpose of the present work is to study linear perturbations of the modified Cartesian flat  cosmological model. 
We find the prediction of our model for the CMB and the growth of structure in the linear regime. We use the available data to constrain the parameters of our proposed model. Furthermore, we discuss the implications of this model in connection with the Hubble constant tension \cite{DiValentino:2021izs}.

The structure of this work is as follows: In Section \ref{sec2}, we review the theoretical background of the local limit of nonlocal gravity first proposed in \cite{Tabatabaei:2022tbq}. We briefly describe the teleparallel extension of general relativity and present the field equations of nonlocal gravity (NLG) and its local limit. The modified Cartesian flat model is a particular solution of the field equations of the local limit of NLG and its properties are studied in Section \ref{sec3}.  In Section \ref{sec4}, linear perturbations of this background cosmological model are studied in the Newtonian gauge. We show in detail the scalar-vector-tensor decomposition of the perturbations and also write the perturbed field and conservation equations in the synchronous gauge to check the model with the cosmic microwave background (CMB) data. This is an essential step as the \texttt{CAMB} code, which is a Boltzmann equation solver based on the gravity model, works in the synchronous gauge \cite{camb}.
In Section \ref{sec5}, we present our results and find the best-fitting parameters of our model. The data we employ in this section involve the cosmic microwave background (CMB) temperature anisotropy angular power spectrum, nearby Supernovae type Ia (SNeIa) and baryon acoustic oscillation (BAO). Finally, Section \ref{sec6} is devoted to a discussion of our results.
% @@@@@@@@@@@@@@@@@@@
% @@@@@@@@@@@@@@@@@@@
% @@@@@@@@@@@@@@@@@@@
% @@@@@@@@@@@@@@@@@@@
\section{Local Limit of Nonlocal gravity: Theoretical Background}
\label{sec2}
Nonlocal gravity (NLG) is a classical nonlocal generalization of general relativity (GR) based on its teleparallel equivalent~\cite{Hehl:2008eu, Hehl:2009es}. For a comprehensive account of NLG, see~\cite{BMB}. The teleparallel equivalent of general relativity (TEGR) is a tetrad theory; that is, the framework of GR is supplemented with a latticework of preferred orthonormal tetrad frame fields that are \emph{parallel} via the Weitzenb\"ock connection. To see how this comes about, consider a gravitational field with spacetime metric
\begin{equation}\label{I1}
ds^2 = g_{\mu \nu}\,dx^\mu \, dx^\nu\,, 
\end{equation}
where we have a preferred set of observers with adapted orthonormal  tetrads $e^\mu{}_{\hat {\alpha}}(x)$, namely, 
\begin{equation}\label{I2}
 g_{\mu \nu}(x) \, e^\mu{}_{\hat {\alpha}}(x)\, e^\nu{}_{\hat {\beta}}(x)= \eta_{\hat {\alpha} \hat  {\beta}}\,.
\end{equation}
In our convention, Greek indices run from 0 to 3, while Latin indices run from 1 to 3; moreover, the signature of the metric is +2 and $\eta_{\alpha \beta}$ is the Minkowski metric tensor given by diag$(-1,1,1,1)$. We use units such that $c = 1$, unless specified otherwise. Furthermore, hatted indices specify the tetrad axes in the local tangent space, while indices without hats are normal spacetime indices.

It is natural to use the preferred tetrad frame field to define the \emph{Weitzenb\"ock connection}~\cite{We}
\begin{equation}\label{I3}
\Gamma^\mu_{\alpha \beta}=e^\mu{}_{\hat{\rho}}~\partial_\alpha\,e_\beta{}^{\hat{\rho}}\,,
\end{equation}
which is nonsymmetric and curvature free. Let $\nabla$ denote covariant differentiation with respect to the Weitzenb\"ock connection; then, $\nabla_\nu\,e_\mu{}^{\hat{\alpha}}=0$, which means that the spacetime becomes a  parallelizable manifold via the Weitzenb\"ock connection. In this framework of teleparallelism, distant vectors are  parallel if they have the same components with respect to the local preferred frame field~\cite{Itin:2018dru, Maluf:2013gaa, Aldrovandi:2013wha}. 
The torsion tensor of the Weitzenb\"ock connection is given by
\begin{equation}\label{I4}
 C_{\mu \nu}{}^{\alpha}=\Gamma^{\alpha}_{\mu \nu}-\Gamma^{\alpha}_{\nu \mu}=e^\alpha{}_{\hat{\beta}}\Big(\partial_{\mu}e_{\nu}{}^{\hat{\beta}}-\partial_{\nu}e_{\mu}{}^{\hat{\beta}}\Big)\,,
\end{equation}
which represents the gravitational field in the context of teleparallelism. It is interesting to compare and contrast the Weitzenb\"ock connection with the Levi-Civita connection of GR given by the Christoffel symbols
\begin{equation}\label{I5}
{^0}\Gamma^\mu_{\alpha \beta}= \frac{1}{2} g^{\mu \nu} (g_{\nu \alpha,\beta}+g_{\nu \beta,\alpha}-g_{\alpha \beta,\nu})\,.
\end{equation}
This torsion-free connection has Riemannian curvature ${^0}R_{\alpha \beta \gamma \delta}$  that in GR represents the gravitational field.  We employ a left superscript ``0" to designate geometric quantities directly related to the Levi-Civita connection. Einstein's gravitational field equations are then given by
\begin{equation}\label{I6}
{^0}G_{\mu \nu} + \Lambda\, g_{\mu \nu}=\kappa\,T_{\mu \nu}\,, 
 \end{equation}
 where $T_{\mu \nu}$ is the symmetric energy-momentum tensor of matter, $\kappa:=8 \pi G/c^4$ and the Einstein tensor ${^0}G_{\mu \nu}$ is defined as  
\begin{equation}\label{I7}
 {^0}G_{\mu \nu} := {^0}R_{\mu \nu}-\frac{1}{2} g_{\mu \nu}\,{^0}R\,,
 \end{equation} 
where ${^0}R_{\mu \nu} = {^0}R^{\alpha}{}_{\mu \alpha \nu}$ is the Ricci tensor and ${^0}R = {^0}R^{\mu}{}_{\mu}$ is the scalar curvature.

On a given manifold,  the difference between two connections is always a tensor. Indeed,  the \emph{contorsion} tensor is defined by
\begin{equation}\label{I8}
K_{\mu \nu}{}^\alpha= {^0} \Gamma^\alpha_{\mu \nu} - \Gamma^\alpha_{\mu \nu}\,.
\end{equation}
Let us note that $\nabla_\nu\, g_{\alpha \beta}=0$, which follows from the orthonormality of the preferred tetrad frame field. Therefore, the Weitzenb\"ock connection is metric compatible; indeed, we have two metric-compatible connections in our extended GR framework. Moreover, the metric-compatibility of the Weitzenb\"ock connection can be used to relate contorsion to torsion, namely, 
\begin{equation}\label{I9}
K_{\mu \nu \rho} = \frac{1}{2}\, (C_{\mu \rho \nu}+C_{\nu \rho \mu}-C_{\mu \nu \rho})\,.
\end{equation}
The torsion tensor is antisymmetric in its first two indices; hence, the contorsion tensor is antisymmetric in its last two indices. In this extended GR framework, the torsion of the Weitzenb\"ock connection, $C_{\mu \nu \rho}$, and the curvature of the Levi-Civita connection,  ${^0}R_{\mu \nu \rho \sigma}$, are complementary features of the gravitational field. 

The Levi-Civita connection is the sum of the contorsion tensor plus the Weitzenb\"ock connection; therefore, one can reformulate GR in terms of the torsion tensor. The result is TEGR, the teleparallel equivalent of GR,  which is the subject of the next subsection. 
% &&&&&&&&&&&&&&&&&&&&&&&&&&&&&&&
% &&&&&&&&&&&&&&&&&&&&&&&&&&&&&&&
\subsection{Teleparallel Equivalent of General Relativity}
Within the framework of teleparallelism, one can write the Einstein tensor as~\cite{BMB}
\begin{eqnarray}\label{G1}
 {^0}G_{\mu \nu}=\frac{\kappa}{\sqrt{-g}}\Big[e_\mu{}^{\hat{\gamma}}\,g_{\nu \alpha}\, \frac{\partial}{\partial x^\beta}\,\mathfrak{H}^{\alpha \beta}{}_{\hat{\gamma}}
-\Big(C_{\mu}{}^{\rho \sigma}\,\mathfrak{H}_{\nu \rho \sigma}
-\frac{1}{4}\,g_{\mu \nu}\,C^{\alpha \beta \gamma}\,\mathfrak{H}_{\alpha \beta \gamma}\Big) \Big]\,,
\end{eqnarray}
where the auxiliary torsion field $\mathfrak{H}_{\mu \nu \rho}$ is simply proportional to the auxiliary torsion tensor $\mathfrak{C}_{\mu \nu \rho}$ by 
\begin{equation}\label{G2}
\mathfrak{H}_{\mu \nu \rho}:= \frac{\sqrt{-g}}{\kappa}\,\mathfrak{C}_{\mu \nu \rho}\,.
\end{equation}
Here, as in GR,  $g:=\det(g_{\mu \nu})$ and $\sqrt{-g}=\det(e_{\mu}{}^{\hat{\alpha}})$. The auxiliary torsion tensor is antisymmetric in its first two indices, just like the torsion tensor, and is defined by
\begin{equation}\label{G3}
\mathfrak{C}_{\alpha \beta \gamma} :=C_\alpha\, g_{\beta \gamma} - C_\beta \,g_{\alpha \gamma}+K_{\gamma \alpha \beta}\,,
\end{equation}
where $C_\mu$,
\begin{equation}\label{G4}
C_\mu :=C^{\alpha}{}_{\mu \alpha} = - C_{\mu}{}^{\alpha}{}_{\alpha}\,,
\end{equation}
is the torsion vector. 

It is now possible to write Einstein's field equation~\eqref{I6} in the form
\begin{equation}\label{G5}
 \frac{\partial}{\partial x^\nu}\,\mathfrak{H}^{\mu \nu}{}_{\hat{\alpha}}+\frac{\sqrt{-g}}{\kappa}\,\Lambda\,e^\mu{}_{\hat{\alpha}} =\sqrt{-g}\,(T_{\hat{\alpha}}{}^\mu + \mathbb{T}_{\hat{\alpha}}{}^\mu)\,,
\end{equation}
which is therefore the TEGR field equation.  Here,  $\mathbb{T}_{\mu \nu}$,  
\begin{equation}\label{G6}
\sqrt{-g}\,\mathbb{T}_{\mu \nu} :=C_{\mu \rho \sigma}\, \mathfrak{H}_{\nu}{}^{\rho \sigma}-\frac 14  g_{\mu \nu}\,C_{\rho \sigma \delta}\,\mathfrak{H}^{\rho \sigma \delta}\,,
\end{equation}
is the trace-free energy-momentum tensor of the gravitational field. From the TEGR field Eq.~\eqref{G5}, we can derive the law of conservation of total energy-momentum tensor; that is, 
 \begin{equation}\label{G7}
\frac{\partial}{\partial x^\mu}\,\Big[\sqrt{-g}\,(T_{\hat{\alpha}}{}^\mu + \mathbb{T}_{\hat{\alpha}}{}^\mu -\frac{\Lambda}{\kappa}\,e^\mu{}_{\hat{\alpha}})\Big]=0\,,
 \end{equation}
which follows from taking the partial derivative $\partial/\partial x^\mu$ of Eq.~\eqref{G5} and using the antisymmetry of $\mathfrak{H}^{\mu \nu}{}_{\hat{\alpha}}$ in its first two indices. 

It can be shown that TEGR is the gauge theory of the 4-parameter Abelian group of spacetime translations~\cite{Cho, BlHe}. This nonlinear theory is in some ways similar to  Maxwell's electrodynamics in a medium~\cite{HeOb}. To illustrate this point, consider the torsion tensor in the form
\begin{equation}\label{G8}
 C_{\mu \nu}{}^{\hat{\alpha}}=e_\rho{}^{\hat{\alpha}}C_{\mu \nu}{}^{\rho}= \partial_{\mu}e_{\nu}{}^{\hat{\alpha}}-\partial_{\nu}e_{\mu}{}^{\hat{\alpha}}\,,
\end{equation}
where we have an analogue of the electromagnetic field strength $(\mathbf{E},  \mathbf{B})\mapsto F_{\mu \nu} = \partial_\mu A_\nu - \partial_\nu A_\mu$ defined in terms of the vector potential $ A_\mu = e_{\mu}{}^{\hat{\alpha}}$ for each ${\hat{\alpha}}={\hat{0}}, {\hat{1}}, {\hat{2}}, {\hat{3}}$. In the same way, the electromagnetic excitation $(\mathbf{D},  \mathbf{H}) \mapsto H_{\mu \nu}$ is similar to $\mathfrak{H}_{\mu \nu}{}^{\hat \alpha}$, while the TEGR field equation~\eqref{G5} is similar to Maxwell's equation $\partial_{\nu}H^{\mu \nu} = 4 \pi \,J^{\mu}$, where $J^\mu$ is the free electric current 4-vector. In Maxwell's theory, the constitutive relation connects the excitation $H_{\mu \nu}$ to 
the field strength $F_{\mu \nu}$; similarly, we have in TEGR the local constitutive relation
\begin{equation}\label{G9}
\mathfrak{H}_{\alpha \beta \gamma} =  \frac{\sqrt{-g}}{\kappa}\,\mathfrak{C}_{\alpha \beta \gamma} = \frac{\sqrt{-g}}{\kappa}\,\left[\frac{1}{2}(C_{\gamma \beta \alpha } + C_{\alpha \beta \gamma} -C_{\gamma \alpha \beta}) +C_\alpha\, g_{\beta \gamma} - C_\beta \,g_{\alpha \gamma}\right]\,,
\end{equation} 
since it connects $\mathfrak{H}_{\alpha \beta \gamma}$ to $C_{\alpha \beta \gamma}$. 

The physical link established here between TEGR and the electrodynamics of media will be of basic importance as we seek to modify GR and hence TEGR; that is, we can use our intuition regarding electrodynamics to guide us in  the teleparallel extension of GR. In particular,  Maxwell's basic equations remain the same when we describe the electrodynamics of different media, only the constitutive relation changes from one medium to another. We follow this same approach as we seek to extend TEGR; that is, we keep the TEGR field equation~\eqref{G5}, but modify the constitutive relation~\eqref{G9}. 
% &&&&&&&&&&&&&&&&&&&&&&&&&&&&&&&
% &&&&&&&&&&&&&&&&&&&&&&&&&&&&&&&
\subsection{Teleparallel Extension of General Relativity}

As explained in the previous section,  we change the constitutive relation~\eqref{G9} in order to modify GR. A simple generic way to do this would be to introduce  a tensor $N_{\mu \nu \rho} = - N_{\nu \mu \rho}$  related to the torsion tensor in Eq.~\eqref{G9}; that is, the new modified constitutive relation is given by
\begin{equation}\label{T1}
\mathcal{H}_{\mu \nu \rho} = \frac{\sqrt{-g}}{\kappa}(\mathfrak{C}_{\mu \nu \rho}+ N_{\mu \nu \rho})\,,
\end{equation} 
where we recover TEGR for $N_{\mu \nu \rho} \to  0$ and $\mathcal{H}_{\mu \nu \rho} \to \mathfrak{H}_{\alpha \beta \gamma}$.  For the modified field equation, we simply replace $\mathfrak{H}_{\mu \nu \rho}$ in Eqs.~\eqref{G5}--\eqref{G6} by $\mathcal{H}_{\mu \nu \rho}$, namely, 
\begin{equation}\label{T2}
 \frac{\partial}{\partial x^\nu}\,\mathcal{H}^{\mu \nu}{}_{\hat{\alpha}}+\frac{\sqrt{-g}}{\kappa}\,\Lambda\,e^\mu{}_{\hat{\alpha}} =\sqrt{-g}\,(T_{\hat{\alpha}}{}^\mu + \mathcal{T}_{\hat{\alpha}}{}^\mu)\,,
\end{equation}
where $\mathcal{T}_{\mu \nu}$ is the new trace-free energy-momentum tensor of the gravitational field given by
\begin{equation}\label{T3}
\kappa\,\mathcal{T}_{\mu \nu} = \kappa\,\mathbb{T}_{\mu \nu} + Q_{\mu \nu}\,.
\end{equation}
Here, $Q_{\mu \nu}$ is a trace-free tensor proportional to $N_{\mu \nu \rho}$,
\begin{equation}\label{T4}
Q_{\mu \nu} := C_{\mu \rho \sigma} N_{\nu}{}^{\rho \sigma}-\frac 14\, g_{\mu \nu}\,C_{ \delta \rho \sigma}N^{\delta \rho \sigma}\,.
\end{equation} 
The conservation law of the total energy-momentum now takes the form
\begin{equation}\label{T5}
\frac{\partial}{\partial x^\mu}\,\Big[\sqrt{-g}\,(T_{\hat{\alpha}}{}^\mu + \mathcal{T}_{\hat{\alpha}}{}^\mu -\frac{\Lambda}{\kappa}\,e^\mu{}_{\hat{\alpha}})\Big]=0\,.
 \end{equation}
 We have presented here a simple constitutive framework for the teleparallel extension of GR. Einstein's GR is based on the metric tensor $g_{\mu \nu}$, which is basically determined by the GR field equations. On the other hand, the modified field equations  should determine the 16 components of the fundamental orthonormal tetrad  frame field $e^\mu{}_{\hat{\alpha}}$. Of these, 10 fix the components of the metric tensor $g_{\mu \nu}$ via the orthonormality condition, while the other 6 are local Lorentz degrees of freedom representing boosts and rotations. 
 
 It is interesting to derive the modified Einstein field equations in this general case. Using Eqs.~\eqref{G9} and~\eqref{T1}, we find
 \begin{equation}\label{T6}
\mathfrak{H}_{\mu \nu \rho} = \mathcal{H}_{\mu \nu \rho} - \frac{\sqrt{-g}}{\kappa} N_{\mu \nu \rho}\,,
\end{equation}
which we plug into the Einstein tensor~\eqref{G1}. Employing Eq.~\eqref{T2}, the 16 modified GR field equations take the form
 \begin{equation}\label{T7}
^{0}G_{\mu \nu} + \Lambda g_{\mu \nu} = \kappa T_{\mu \nu}   + \mathbb{R}_{\mu \nu}\,, \qquad   \mathbb{R}_{\mu \nu} := Q_{\mu \nu} -  \mathcal{N}_{\mu \nu}\,,
\end{equation}
where the tensor $\mathcal{N}_{\mu \nu}$ is defined by
\begin{equation}\label{T8}
\mathcal{N}_{\mu \nu} := g_{\nu \alpha} e_\mu{}^{\hat{\gamma}} \frac{1}{\sqrt{-g}} \frac{\partial}{\partial x^\beta}\,(\sqrt{-g}N^{\alpha \beta}{}_{\hat{\gamma}})\,.
\end{equation} 
The modified GR field equations can be naturally split into symmetric and antisymmetric parts. In fact, we have the 10 modified Einstein equations 
\begin{equation}\label{T9}
^{0}G_{\mu \nu} +   \Lambda g_{\mu \nu}  = \kappa T_{\mu \nu} + Q_{(\mu \nu)}  - \mathcal{N}_{(\mu \nu)}\,
\end{equation}
and the 6 constraint equations
\begin{equation}\label{T10}
Q_{[\mu \nu]} = \mathcal{N}_{[\mu \nu]}\,.
\end{equation}
To go forward, we need to specify the exact relationship between $N_{\mu \nu \rho}$ and $C_{\mu \nu \rho}$.  By using this approach, a nonlocal relation has thus far led to nonlocal gravity (NLG) theory that is patterned after the nonlocal electrodynamics of media~\cite{Hehl:2008eu, Hehl:2009es, BMB}. However,   the local limit of this nonlocal relation is the main focus of the present investigation. 
% !!!!!!!!!!!!!!!!!!!!!!!!!!!!!!!!!!!!!!!!!!!!!!!!!!!!!!!!!!
\subsubsection{Nonlocal Gravity (NLG)}
Let us assume that on the basis of measurements performed by the preferred observers, the components of $N_{\mu \nu \rho}$ are related to those of the torsion tensor $C_{\mu \nu \rho}$ via~\cite{Puetzfeld:2019wwo, Mashhoon:2022ynk}
\begin{equation}\label{N1}
N_{\hat \mu \hat \nu \hat \rho}(x) =  \int  \mathcal{K}(x, x')\,X_{\hat \mu  \hat \nu  \hat \rho }(x') \sqrt{-g(x')}\, d^4x' \,,
\end{equation} 
where the scalar $\mathcal{K}(x, x')$ is the fundamental causal kernel of NLG and
\begin{equation}\label{N2}
X_{\hat \mu \hat \nu \hat \rho}= \mathfrak{C}_{\hat \mu \hat \nu \hat \rho}+ \check{p}\,(\check{C}_{\hat \mu}\, \eta_{\hat \nu \hat \rho}-\check{C}_{\hat \nu}\, \eta_{\hat \mu \hat \rho})\,.
\end{equation}
A detailed explanation for the choice of $X_{\mu \nu \rho}$ is contained in~\cite{BMB}, which should be consulted for an extensive treatment of the physical basis of NLG. In Eq.~\eqref{N2}, $\check{C}_\mu$ is the torsion pseudovector defined via the Levi-Civita tensor $\epsilon_{\alpha \beta \gamma \delta}$ by
\begin{equation}\label{N3}
\check{C}_\mu :=\frac{1}{3!} C^{\alpha \beta \gamma}\,\epsilon_{\alpha \beta \gamma \mu}\,,
\end{equation}
while $\check{p}\ne 0$ is a constant dimensionless parameter. 

The kernel $\mathcal{K}(x, x')$ must be ultimately determined by observation. In this connection, linearized NLG has been treated in detail in~\cite{BMB} and the Newtonian regime of the theory has been compared with observational data~\cite{Rahvar:2014yta, Chicone:2015coa, Roshan:2021ljs, Roshan:2022zov, Roshan:2022ypk}. Within the nonlinear regime of NLG, it has been possible to show that de Sitter spacetime is not an exact solution~\cite{Mashhoon:2022ynk}; indeed, the only known exact solution of NLG is the trivial solution, namely, we find the Minkowski spacetime in the absence of gravitation.  

% !!!!!!!!!!!!!!!!!!!!!!!!!!!!!!!!!!!!!!!!!!!!!!!!!!!!!!!!!!
\subsubsection{Local Limit of Nonlocal Gravity}
In dealing with the nonlocal electrodynamics of media, it is often convenient to introduce local quantities for the polarizability and magnetizability of the medium and write the corresponding local constitutive relations as $\mathbf{D} = \epsilon (x)\mathbf{E}$ and $\mathbf{B} = \mu (x)\mathbf{H}$, where the electric permittivity $\epsilon (x)$ and magnetic permeability $\mu(x)$ are characteristics of the medium. For NLG, a similar procedure would involve the 4D Dirac delta function; that is, 
\begin{equation}\label{N4}
\mathcal{K}(x, x') := \frac{S(x)}{\sqrt{-g(x)}}\,\delta(x-x')\,.
\end{equation}
Here, $S(x)$ is a dimensionless scalar \emph{susceptibility} that is characteristic of the spacetime under consideration. This function, as with the nonlocal kernel,  must ultimately be determined on the basis of observational data.  

The constitutive relation in this case is given by 
\begin{equation}\label{N5}
\mathcal{H}_{\mu \nu \rho} = \frac{\sqrt{-g}}{\kappa}[(1+S)\,\mathfrak{C}_{\mu \nu \rho}+ S\,\check{p}\,(\check{C}_\mu\, g_{\nu \rho}-\check{C}_\nu\, g_{\mu \rho})]\,,
\end{equation}
which reduces to the case of TEGR for $S=0$. If $S(x) \ne 0$, we have a local generalization of GR with a new scalar function $S(x)$. Moreover, we must have $1+S > 0$; otherwise, GR will not be a limiting case of the new theory. In conformity with the electrodynamics of media, $S(x)$ must be compatible with the nature of the background spacetime.

In recent papers, we have employed the local limit of NLG to study FLRW cosmological models~\cite{Tabatabaei:2022tbq, Tabatabaei:2023qxw}. The modified theory under consideration is a tetrad theory; therefore, we must pay particular attention to the tetrad frame of the model from which the FLRW metric is obtained via orthonormality.  The spatially homogeneous and isotropic FLRW models are dynamic and time dependent; hence, we expect on physical grounds that $dS/dt \ne 0$. It turns out that this requirement is satisfied only in the case of the modified Cartesian flat model~\cite{Tabatabaei:2022tbq}. That is, the adapted orthonormal tetrad frame of the standard preferred comoving observers should be along the Cartesian coordinate axes of the background spatially Euclidean FLRW spacetime. 

The rest of this paper is devoted to the cosmological implications of the modified Cartesian flat model. We recall the main properties of the model in the following section and in Section \ref{sec4} we describe the corresponding linear perturbations in detail.  We note that the cosmological implications of teleparallel theories of gravity have been the subject of recent investigations~\cite{Bahamonde:2021gfp, Heisenberg:2022mbo}.
% @@@@@@@@@@@@@@@@@@@
% @@@@@@@@@@@@@@@@@@@
% @@@@@@@@@@@@@@@@@@@
% @@@@@@@@@@@@@@@@@@@
\section{Modified Cartesian Flat Cosmological  Model}
\label{sec3}
Consider a spatially flat FLRW model with preferred observers that are spatially at rest and have adapted orthonormal tetrad axes that are parallel to the Cartesian coordinate directions, namely, 
\begin{equation}\label{M1}
e^\mu{}_{\hat{\alpha}}= \frac{1}{a(\eta)}\,\delta^\mu_{\alpha}\,, \qquad     e_\mu{}^{\hat{\alpha}}= a(\eta)\,\delta_\mu^{\alpha}\,.
\end{equation}
Here $a(\eta)$ is the scale factor and $\eta$ is the standard \emph{conformal time} $x^0 = \eta$ such that $x^\mu = (\eta, x^i)$. The corresponding metric is thus conformally flat, 
\begin{equation}\label{M2}
ds^2= a^2(\eta) \, \eta_{\mu \nu}\, dx^\mu\,dx^\nu\,. 
\end{equation}
The conformal time $\eta$ is related to \emph{cosmic time} $t$ via $dt = a(\eta) d\eta$. Using standard conventions, we can write
\begin{equation}\label{M3}
\dot a := \frac{da}{dt} = \frac{1}{a(\eta)}\,\frac{da(\eta)}{d\eta} := \frac{a'}{a} = \mathcal{H}(\eta) = a H(t)\,, 
\end{equation}
where a prime indicates differentiation with respect to conformal time and  $H$ is the Hubble parameter. 

The torsion tensor for our Cartesian flat model is given by
\begin{equation}\label{M4}
 C_{\alpha \beta \gamma}= a a' (\delta^{0}_{\alpha}\, \eta_{\beta \gamma}- \delta^{0}_{\beta }\, \eta_{\alpha \gamma})\,.
\end{equation}
We find the torsion vector, the contorsion tensor and the auxiliary torsion tensor are 
\begin{equation}\label{M5}
 C_{\alpha}= -3 \frac{a'}{a}\,\delta_{\alpha 0}\,, \qquad K_{\alpha \beta \gamma} = C_{\beta \gamma \alpha}\,, \qquad  \mathfrak{C}_{\alpha \beta \gamma}=-2\,C_{\alpha \beta \gamma}\,.
\end{equation}
Moreover, $N_{\alpha \beta \gamma} = -2S(\eta)\,C_{\alpha \beta \gamma}$, since $\check{C}_\alpha=0$ in this case.

The Einstein tensor for our conformally flat spacetime is diagonal; its components are given by $^0G_{0i} = 0$ and
\begin{equation}\label{M6}
^0G_{00}= 3 \left(\frac{a'}{a}\right)^2\,, \qquad ^0G_{ij}= - \left[2\frac{a''}{a} - \left(\frac{a'}{a}\right)^2\right]\,\delta_{ij}\,.
\end{equation}
 The source of the gravitational field is a perfect fluid of energy density $\rho(\eta)$, pressure $P(\eta)$ and energy-momentum tensor
\begin{equation}\label{M7}
T_{\mu \nu}=\rho\,u_\mu\,u_\nu+P\,(g_{\mu \nu}+u_\mu\,u_\nu)\,,
\end{equation}
where $u^\mu$ is the 4-velocity vector of the perfect fluid. We assume, as in the standard models, that the particles of the fluid are comoving with the preferred observers and are thus spatially at rest, namely, 
\begin{equation}\label{M8}
 u^\mu = e^{\mu}{}_{\hat 0} = \frac{1}{a(\eta)}\delta^\mu_0\,,  \qquad u_\mu = a(\eta)\eta_{\mu 0}\,.
\end{equation} 
 
As demonstrated before~\cite{Tabatabaei:2022tbq, Tabatabaei:2023qxw}, field Eqs.~\eqref{T9} and~\eqref{T10} are satisfied in this case with $S(\eta)$. More specifically, we find 
\begin{equation}\label{M9}
 \frac{3}{a^2}(1+S) \left( \frac{a'}{a}\right)^2 = \Lambda + 8\pi G \rho\,, 
\end{equation} 
\begin{equation}\label{M10}
 2(1+S) \left( \frac{a'}{a}\right)' + (1+S) \left( \frac{a'}{a}\right)^2 = a^2(\Lambda - 8\pi G P) - 2 \frac{dS}{d\eta}\frac{a'}{a}\,, 
\end{equation} 
respectively. As expected,  the standard GR results are recovered when $S=0$. Furthermore, it has been shown that the modified Cartesian flat model is inconsistent with the existence of the cosmological 
constant $\Lambda$, just as in NLG; therefore, we must resort to dynamic dark energy in order to account for the accelerated expansion of the universe~\cite{Tabatabaei:2022tbq, Tabatabaei:2023qxw}. Employing $\mathcal{H} = a'/a$, the unperturbed equations for our modified model become
\begin{equation}\label{M11}
 3 (1+S) \mathcal{H}^2 =  \kappa a^2 \rho\,, 
\end{equation} 
\begin{equation}\label{M12}
 2(1+S) \mathcal{H}' + (1+S)  \mathcal{H}^2 = - \kappa a^2 P - 2 \frac{dS}{d\eta}\mathcal{H}\,. 
\end{equation} 

The main result of our approach is the standard model of cosmology supplemented with a function $S(\eta)$. At late times, de Sitter spacetime is not a solution of the modified model for $dS/d\eta \ne 0$; hence, the cosmological constant must be replaced with dynamic dark energy~\cite{Tabatabaei:2023qxw}. The relationship between this modified model and the standard $\Lambda$CDM model is explored in Appendix A.  

Nonlocal gravity (NLG) contains characteristic spatial and temporal scales associated with nonlocality that affect the nature of the gravitational field on large scales~\cite{Bouche:2022jts}. These scales are seemingly expected to disappear in the local limit of NLG. On the other hand, the kernel of NLG in the local limit takes shape as the gravitational susceptibility function $S(x)$ much like the electric permittivity $\epsilon(x)$ and magnetic permeability $\mu(x)$ in the case of the electrodynamics of media. We expect that $S(x)$ represents an intrinsic property of spacetime and captures something of the full nonlocal characteristics of the gravitational field. Indeed, the late-time behavior of the modified Cartesian flat cosmological model is dominated by dynamic dark energy since the existence of the cosmological constant is ruled out in this modified local model just as de Sitter spacetime is excluded as an exact solution of NLG~\cite{Mashhoon:2022ynk, Tabatabaei:2023qxw}. Furthermore, $S$ is in effect a function of the scale parameter $a$~\cite{Tabatabaei:2022tbq}. For the sake of definiteness, we assume in Section 5 a simple form for this relation, namely, $S$ is proportional to $a$ raised to a positive power; therefore, $dS/d\eta$ is indeed proportional to the Hubble parameter $H$, which is the basic scale of the model under consideration here.

%@@@@@@@@@@@@@@@@@@@@
%@@@@@@@@@@@@@@@@@@@@
%@@@@@@@@@@@@@@@@@@@@
%@@@@@@@@@@@@@@@@@@@@
\section{Cosmological Perturbations}
\label{sec4}

In this section, we treat the linear perturbation theory of our modified Cartesian flat cosmological model.
In the standard GR case, one chooses a perturbed FLRW metric in a specific gauge and finds the Einstein equations in terms of the first-order perturbed metric components. In the absence of a cosmological constant, GR implies $\delta G_{\mu\nu} = \kappa \delta T_{\mu\nu}$,  which results in the relativistic Poisson equation~\cite{Amendola:2015ksp}. On the other hand, for the local limit of nonlocal gravity (NLG), we perturb the tetrad frame field. The local theory under consideration here in general allows the possibility of parity-violating solutions proportional to the constant dimensionless parameter $\check{p}$. To simplify matters, in this first treatment of cosmological perturbations, we simply ignore this possibility by working with the local limit of NLG with $\check{p}=0$ in Eq.~\eqref{N5}.

In our perturbation analysis, the perturbed susceptibility function can be a function of position as well as time and can be written as 
\begin{equation}\label{Q1}
S (\eta,\mathbf{x}) = \bar{S}(\eta) + \delta S(\eta,\mathbf{x})\,,  
\end{equation} 
where $\bar{S}(\eta)$ is the unperturbed value of this function. Henceforth, we use a ``bar" to indicate background unperturbed quantities. It proves useful to employ Fourier transformation such that 
\begin{equation}\label{Q2}
\widetilde{\delta S}(\eta,\mathbf{k}) = \int d^3x \,e^{-i \mathbf{k} \cdot \mathbf{x}} \,\delta S(\eta, \mathbf{x})\,.
\end{equation}
The resulting Fourier transform is a complex function; as usual, however, only its real part has physical significance in our linear perturbation scheme. 
Henceforth, we use a ``tilde" to indicate quantities in Fourier space.
In the following subsections, we investigate the tetrad perturbations and the corresponding perturbed field equations.

% @@@@@@@@@@@@@@@@@@@@@
% @@@@@@@@@@@@@@@@@@@@@

\subsection{Perturbation of Tetrads and Energy-Momentum Tensor}

We write the perturbed tetrad frame field for our modified TEGR cosmological model in the form 
\begin{equation}\label{P1}
e_{\mu}{}^{ \hat{\alpha}}(x)  = a(\eta)[\delta_{\mu}^{\alpha}+\psi_{\mu}{}^{\hat{\alpha}}(x)]\,, \qquad e^{\mu}{}_{ \hat{\alpha}}(x)  = a^{-1}(\eta)[\delta^{\mu}_{\alpha}-\psi^{\mu}{}_{\hat{\alpha}}(x)]\,,
\end{equation} 
where $\psi_{\mu}{}^{\hat{\alpha}}$ and $\psi^{\mu}{}_{\hat{\alpha}}$  are treated to linear order away from the Minkowski background and, accordingly,  at this order the distinction between spacetime and tetrad indices can henceforward be ignored. 
It follows from the tetrad orthonormality condition, $e_{\mu}{}^{ \hat{\alpha}}e_{\nu}{}^{ \hat{\beta}}\eta_{\hat{\alpha} \hat{\beta}} = g_{\mu \nu}$,  
that the metric tensor of the perturbed spacetime is given by
\begin{equation}\label{P2}
g_{\mu \nu} = a^2(\eta)(\eta_{\mu \nu} + h_{\mu \nu})\,, \qquad h_{\mu \nu}:=2\psi_{(\mu \nu)}\,.
\end{equation}
It remains to express $\psi_{\mu \nu}(\eta, \mathbf{x})$ in suitable form. 

We employ the scalar-vector-tensor decomposition of cosmological perturbations~\cite{Golovnev:2018wbh, Bahamonde:2020lsm, Hohmann:2020vcv} and write 
\begin{equation}\label{P3}
\psi_{00} := \psi\,, \qquad \psi_{i0} := -(\partial_i \varpi + \gamma_i)\,, \qquad \psi_{0i} := \partial_i \zeta + v_i\,, 
\end{equation}
\begin{equation} \label{P4}
\psi_{ij} := \delta_{ij}\, \phi + \partial_i\,\partial_j \varsigma + \epsilon_{ijk}\,\partial_k \mathfrak{s} + \partial_j c_i+\epsilon_{ijk}\,\mathfrak{v}^k+\frac{1}{2}\,\chi_{ij}\,,
\end{equation}
where $\psi$, $\varpi$, $\zeta$, $\phi$, and $\varsigma$ are scalar functions, while $\gamma_i$, $v_i$, and $c_i$ are all divergenceless vector functions. Moreover, $\mathfrak{s}$ and $\mathfrak{v}_i$  are pseudoscalar  and pseudovector functions, respectively. Finally,  $\chi_{ij}$ is a symmetric and traceless tensor that is also transverse, i.e. $\delta^{ik}\,\partial_k \chi_{ij} = 0$.  With our choice of tetrad functions, the metric is given by
\begin{equation}\label{P5}
g_{00} = -a^2(\eta)(1-2\psi)\,, \quad g_{0i} = a^2(\eta)[\partial_i(\zeta-\varpi)+v_i-\gamma_i)]\,, 
\end{equation}
\begin{equation}\label{P6}
 g_{ij} = a^2(\eta) [(1+2\phi)\delta_{ij}+2\partial_i\partial_j \varsigma + \partial_i\,c_j+\partial_j\,c_i+ \chi_{ij}]\,.
\end{equation}
Except for the pseudoscalar ($\mathfrak{s}$) and pseudovector ($\mathfrak{v}_i$) terms, the other functions appear in the metric. The gauge transformation of the tetrad functions has been discussed in~\cite{Golovnev:2018wbh}. In what follows, we adopt the conformal Newtonian gauge in which
\begin{equation}\label{P7}
\varsigma = 0\,, \qquad  \zeta = \varpi\,, \qquad c_i=0\,.
\end{equation}

For the sake of simplicity, we discarded parity-violating terms in the field equations at the outset. In this connection, it would make sense to ignore $\mathfrak{s}$ and $\mathfrak{v}_i$ in our treatment. Indeed, it will turn out that $\mathfrak{s}$ does not appear in the perturbed field equations at all and, moreover,  $\mathfrak{v}_i$ can be consistently neglected as  $\epsilon_{ijk} \partial_j\mathfrak{v}_k = 0$. 

It is also essential to determine the perturbed energy-momentum tensor~\eqref{M7}. In its unperturbed form,  
\begin{equation}\label{P7a}
\bar{T}_{\mu \nu} = {\rm diag} [a^2(\eta)\,(\bar{\rho}, \bar{P}, \bar{P}, \bar{P})]\,.
\end{equation}
In the perturbed energy-momentum tensor, $T_{\mu\nu} = (\rho+P)u_\mu\,u_\nu + P\,g_{\mu\nu}$, we ignore heat transfer and  viscous shear stresses. The perturbed energy density and pressure are given by $\rho = \bar{\rho} + \delta\rho$ and $P=\bar{P} + \delta{P}$.  Here we define the density contrast parameter as 
\begin{equation}
\delta := \delta\rho / \bar{\rho}\,.
\end{equation}
Moreover, the perturbed 4-velocity vector  $u^\mu$, $u^\mu u_\mu = -1$, takes the form
\begin{equation}\label{P8}
u^{\mu} = \frac{1}{a}(1+\psi,~ ~\mathfrak{u}^i)\,, \qquad u_{\mu} = a(-1 + \psi,~~v_i-\gamma_i + \mathfrak{u}_i)\,,
\end{equation}
where the spatial velocity perturbation $\mathfrak{u}^i$ is treated to linear order and $\mathfrak{u}_i := \delta_{ij}\,\mathfrak{u}^j$. Employing the Helmholtz decomposition, we can write the vector field $\mathfrak{u}$ in terms of its divergence-free $(\mathfrak{u}_{\perp})$ and curl-free $(\mathfrak{u}_{\parallel})$ components, namely, 
\begin{equation}\label{P9}
\mathfrak{u}^i = \mathfrak{u}^i_{\perp} + \mathfrak{u}^i_{\parallel}\,, \qquad \nabla \cdot \mathfrak{u}_{\perp} = 0\,, \qquad \nabla \times  \mathfrak{u}_{\parallel} = 0\,.
\end{equation}
The expansion parameter for the perturbed spatial velocity field is defined by
\begin{equation}\label{P10}
\theta := \partial_i \,\mathfrak{u}^i =  \nabla \cdot \mathfrak{u}_{\parallel}\,.
\end{equation}

Regarding density and pressure, we define $w$, 
\begin{equation}\label{P10a}
w := \frac{\bar{P}}{\bar{\rho}}\,,
\end{equation} 
which is normally assumed to be a constant. Moreover, we need to specify the relation between  $\delta P$ and $\delta\rho$; this relation, in the rest frame of the fluid,  is defined as the sound speed $c_s \geq 0$, 
\begin{equation}\label{P10b}
c^2_s := \frac{dP}{d\rho}|_{\rm{Fluid Rest Frame}}\,,
\end{equation} 
which is a characteristic feature of the fluid under consideration. For dark matter and baryons, we set $c_s=0$.

%@@@@@@@@@@@
%@@@@@@@@@@@
\subsection{Perturbed Field Equations}

In this subsection, we study the perturbed field equations for our modified TEGR cosmological model.
The modified GR field equations can be naturally split into symmetric and antisymmetric parts as in Eqs.~\eqref{T9} and~\eqref{T10}. In fact, we have the 10 modified perturbed Einstein equations 
\begin{equation}\label{T9-p}
\delta\, ^{0}G_{\mu \nu}  = \kappa  \delta T_{\mu \nu} + \delta Q_{(\mu \nu)}  - \delta \mathcal{N}_{(\mu \nu)}\,
\end{equation}
and the 6 perturbed constraint equations
\begin{equation}\label{T10-p}
\delta Q_{[\mu \nu]} = \delta \mathcal{N}_{[\mu \nu]}\,.
\end{equation}

Let us first consider the perturbed symmetric field Eq.~\eqref{T9-p}. The $(00)$ component of this equation reads
\begin{equation}\label{00raw}
-\Delta\phi + 3\mathcal{H}(\phi'+ \mathcal{H}\psi) =\frac{ \kappa a^2 \delta\rho - 3\mathcal{H}^2 \delta S }{2(1+\bar{S})}\,,
\end {equation} 
where $\Delta$ is the Laplace operator.  Next, the three symmetric $(0i)$ components are 
\begin{equation}\label{0iraw1}
\Delta(\gamma_i- v_i)+ 6(1+w)\,\mathcal{H}^2\,(v_i - \gamma_i + \mathfrak{u}_i) +  \ss\,\epsilon_{ijk}\partial_j \mathfrak{v}_k -2\partial_i\left(2\,\mathcal{H}\psi + \mathcal{H}\frac{\delta S}{1+\bar{S}} + 2\phi' + \ss \phi\right) = 0\,.
\end{equation}
Here, we have introduced the notation $\ss := \bar{S}' /(1+\bar{S})$ for the sake of simplicity. Finally, the six symmetric $(ij)$ components are 
\begin{small}
\begin{equation}\label{ijraw-sym}
\begin{split}
 & -4 \delta_{ij} \left[\Phi + \mathcal{H}\, \Psi + \frac{1}{2} \kappa a^2 \frac{\delta P}{1+\bar{S}} - \mathcal{H}\left( 2 \ss+3w\mathcal{H} \right)\frac{\delta S}{1+\bar{S}}  + 2\mathcal{H} \frac{\delta S'}{1+\bar{S}} \right] \\
 & - 2(\partial_i\partial_j - \delta_{ij} \Delta)\left(\ss\,\zeta -\psi + \phi  \right) + \partial_i \mathbb{V}_j + \partial_j \mathbb{V}_i  \\
&+\chi''_{ij}+\left(2\mathcal{H} + \ss\right)\chi'_{ij} - \Delta\chi_{ij} =0\,. 
\end{split}
\end{equation}
\end{small}
Here, $\Phi$, $\Psi$ and $\mathbb{V}_i$ are given by
\begin{equation}\label{P11}
\Phi = \phi'' + \left(2\mathcal{H} + \ss \right) \phi' \,,
\end {equation} 
\begin{equation}\label{P12}
\Psi = \psi' - 3w\mathcal{H} \psi \,,
\end {equation} 
\begin{equation}\label{P13}
\mathbb{V}_i = 2  \mathcal{H} (\gamma_i - v_i) -\ss \, v_i + \gamma'_i - v'_i\,.
\end {equation}
For $S = 0$ in the 10 modified perturbed Einstein equations, we recover the standard  GR results~\cite{Amendola:2015ksp}. 

For the six antisymmetric equations~\eqref{T10-p}, let us first consider the three $[0i]$ relations, namely, 
\begin{equation} \label{eq-0i-asym}
2\mathcal{H}\partial_i \delta S + \bar{S}'\left(\epsilon_{ijk}\partial_j \mathfrak{v}_k - 2\partial_i\phi\right) =0\,. 
\end{equation}
The remaining three  antisymmetric $[ij]$ relations are given by
\begin{equation}\label{P14}
\bar{S}' \,	(\partial_i \gamma_j -\partial_j \gamma_i)	=0\,.
\end{equation}
For $S=0$, the constraint equations all vanish, as expected. 

We now have the 16 perturbation equations of our modified TEGR model. To discuss the physical content of these equations, let us start with the simplest, namely, Eq.~\eqref{P14}. We know that $\bar{S}' \ne 0$ by assumption; therefore, Eq.~\eqref{P14} implies $\gamma_i = \partial_i \xi$, where $\xi$ is a scalar function that satisfies Laplace's equation, $\Delta \xi = 0$, since the vector $\gamma_i$ is divergence-free in our decomposition. We assume all of the functions that appear in our decomposition are sufficiently smooth and bounded as would be required for our cosmological perturbation scheme. A real harmonic function is a smooth function defined on an open subset of the three-dimensional Euclidean space that satisfies Laplace's equation. According to Liouville's theorem, a bounded harmonic function defined over all of the three-dimensional Euclidean space must be constant. Therefore, $\xi$ is a constant and
\begin{equation}\label{P15}
 \gamma_i	=0\,.
\end{equation}

Next, let us take the  divergence of Eq.~\eqref{0iraw1}. We recall that $\gamma_i = 0$ and $v_i$ is divergence free in our decomposition; therefore, 
\begin{equation}\label{possion_raw}
\Delta\left(2\,\mathcal{H}\,  \psi + \mathcal{H} \frac{\delta S}{1+\bar{S}} +2\phi' +\ss \,  \phi\right)= 3(1+w)\,\mathcal{H}^2\, \theta.
\end{equation}
For $S =  0$, we have the standard GR result, i.e.  $\Delta (\phi' + \mathcal{H}\,  \psi) = \kappa a^2 (1+w)\bar{\rho}\,\theta/2$~\cite{Amendola:2015ksp}, where we have used the unperturbed Eq.~\eqref{M11} in this case, namely,  $3 {\mathcal{H}}^2 =  \kappa a^2 \bar{\rho}$.  Relation~\eqref{possion_raw} is a dynamical equation for the scalar degree of freedom in our perturbation scheme. 
It is also possible to extract the vectorial part of Eq.~\eqref{0iraw1} by taking its curl; we find,
\begin{equation}\label{vectorcurl}
\nabla \times \mathbf{U} = 0\,, \qquad U_i = \Delta v_i - 6(1+w)\,\mathcal{H}^2\,(v_i+\mathfrak{u}_i^{\perp}) - \ss\,\epsilon_{ijk}\partial_j \mathfrak{v}_k\,, 
\end{equation}
where we have used $\gamma_i = 0$ and $\nabla \times  \mathfrak{u}_{\parallel} = 0$.  The vectors $v_i$ and $\mathfrak{u}_i^{\perp}$ in $\mathbf{U}$ are divergence free; therefore, the divergence of $\mathbf{U}$ vanishes. On the other hand, it follows from   Eq.~\eqref{vectorcurl} that $\mathbf{U} = \nabla f$ for a scalar field $f$. Thus, $\Delta f = 0$; as before, Liouville's theorem implies that $f$ is constant and $\mathbf{U} = 0$, namely,
\begin{equation}\label{P16}
\Delta v_i - 6(1+w)\,\mathcal{H}^2\,(v_i+\mathfrak{u}_i^{\perp}) - \ss\,\epsilon_{ijk}\partial_j \mathfrak{v}_k = 0\,.
\end{equation}

Let us now return to Eq.~\eqref{ijraw-sym} and define 
\begin{equation}
 \Theta_{ij} := \chi''_{ij}+\left(2\mathcal{H} +\ss\right)\chi'_{ij} - \Delta\chi_{ij} \,,
\end{equation}
\begin{equation}
\mathbb{S}_1 := \ss \, \zeta -\psi + \phi\,
\end{equation}
and
\begin{small}
\begin{equation}
\mathbb{S}_2 := \Phi+ \mathcal{H} \Psi + 4\pi G a^2 \frac{\delta P}{1+\bar{S}} - \mathcal{H}\left( 2\ss+3w\mathcal{H} \right) \frac{\delta S}{1+\bar{S}} + 2\mathcal{H} \frac{\delta S'}{1+\bar{S}}\,.
\end{equation}
\end{small}
Accordingly, Eq.~\eqref{ijraw-sym} becomes:
\begin{equation} \label{eq-svt1}
\mathbb{T}_{ij}  + \partial_i \mathbb{V}_j + \partial_j \mathbb{V}_i  -2 (\partial_i \partial_j - \delta_{ij} \Delta) \mathbb{S}_1 -4 \delta_{ij} \mathbb{S}_2 = 0\,,
 \end{equation}
 where
 \begin{equation}\label{vector_ij_sym}
- \mathbb{V}_i = 2  \mathcal{H} v_i + \ss \,v_i + v'_i\,.
\end{equation}
Here,   $\mathbb{T}_{ij}$ is a symmetric, transverse and traceless tensor; moreover, vector $\mathbb{V}$ is divergence free. We can rewrite  Eq.~\eqref{eq-svt1} as
 \begin{equation} \label{eq-svt2}
\Theta_{ij}  + \partial_i \mathbb{V}_j + \partial_j \mathbb{V}_i -2 (\partial_i \partial_j - \frac{1}{3} \delta_{ij} \Delta) \mathbb{S}_1 - 4\delta_{ij} (\mathbb{S}_2 -\frac{1}{3}\Delta \mathbb{S}_1) = 0\,.
 \end{equation}
Calculating the trace of \eqref{eq-svt2},  we find
 \begin{equation}
\mathbb{S}_2 - \frac{1}{3}\Delta\mathbb{S}_1 = 0\,.
 \end{equation}
We are thus left with
 \begin{equation}\label{Traceless}
\Theta_{ij} + \partial_i \mathbb{V}_j + \partial_j \mathbb{V}_i -2 (\partial_i \partial_j - \frac{1}{3} \delta_{ij} \Delta) \mathbb{S}_1= 0\,.
 \end{equation}
 Taking  the double divergence ($\partial_i \partial_j$) of Eq.~\eqref{Traceless} results in
 \begin{equation}
 \Delta^2 \mathbb{S}_1 = 0\,.
 \end{equation}
Let us assume that the cosmological perturbation is such that $\phi$, $\psi$ and $\zeta$ vanish at spatial infinity; then, applying Liouville's theorem twice results in $\mathbb{S}_1 =  0$ and subsequently $\mathbb{S}_2 = 0$ as well. The remaining equation is
  \begin{equation}
\Theta_{ij} 
+ \partial_i \mathbb{V}_j + \partial_j \mathbb{V}_i= 0\,.
 \end{equation}
Taking the divergence of this equation, one finds
 \begin{equation}
\Delta\mathbb{V}_i= 0\,.
 \end{equation}
Assuming that $v_i$ tends to zero at spatial infinity, Liouville's theorem and Eq.~\eqref{vector_ij_sym} imply $\mathbb{V}_i=0$ and this subsequently results in $\Theta_{ij}=0$.

Finally,  it is possible to decompose Eq.~\eqref{eq-0i-asym} into scalar and pseudo-vectorial parts. Writing down the divergence of  this equation gives $\Delta(\mathcal{H}\delta S -  \bar{S}' \phi ) =0$. It follows from Liouville's theorem that
\begin{equation}\label{scalar_0i_asym}
\mathcal{H}\,\delta S - \bar{S}'\,\phi = 0\,,
\end{equation}
assuming localized perturbations, as before, such that both $\phi$ and $\delta S$ vanish at infinity.
Substituting this result in Eq.~\eqref{eq-0i-asym}, we obtain
\begin{equation}\label{P17}
\epsilon_{ijk}\partial_j \mathfrak{v}_k=0\,,
\end{equation}
since $\bar{S}' \ne 0$ by assumption. The pseudovector $\mathfrak{v}_k$ appears in our perturbation equations only as in Eq.~\eqref{P17}; therefore, we can essentially ignore the pseudovector $\mathfrak{v}_k$ in our subsequent analysis. Furthermore, the pseudoscalar term $\mathfrak{s}$ does not appear in the perturbed field equations and we have the freedom to set it equal to zero.  In connection with parity-violating terms, let us recall that for simplicity we imposed at the outset the requirement that $\check{p}=0$; if we relax this condition, the situation may be different and should be reconsidered.
In the following, we will examine the equations of motion in each sector.

%!!!!!!!!!!!!!!!!!!!!!!!!!!!!!!!!!!!!!!!!!!
\subsubsection{Scalar Sector}

The remaining perturbation equations contain the scalar fields $\phi$, $\psi$, $\zeta$, $\delta S$, $\delta \rho$ and $\delta P$. The first three appear in the perturbed metric and are related by
\begin{equation}\label{phipsirel}
\phi -\psi + \ss\,\zeta=0\,.
\end{equation}
Of these, only $\zeta$ has no dynamical equation for its determination. It is therefore possible to set $\zeta = 0$ in a way that is completely consistent with the remaining perturbations. Hence, $\phi = \psi$. Moreover, it follows from Eq.~\eqref{scalar_0i_asym} that 
\begin{equation}\label{delta_s_sol}
\delta S= \frac{\bar{S}'}{\mathcal{H}}\,\phi\,.
\end{equation}
Using these results, Eq.~\eqref{00raw} turns into 
\begin{equation}\label{R1}
\phi' + \left(-\frac{1}{3\mathcal{H}}\Delta+ \mathcal{H}+\frac{1}{2}\ss\right)\phi = \frac{\kappa\,a^2}{6(1+\bar{S})\mathcal{H}}\delta\rho\,.
\end{equation} 

From $\mathbb{S}_2 = 0$, we find (with $\phi=\psi$), 
\begin{equation}\label{P8.5}
\phi'' + (3\mathcal{H} +\ss) \phi'  - (3w\mathcal{H}^2 + 3w\mathcal{H} \ss + 2 \ss^2) \phi + \frac{1}{2} \kappa a^2 \frac{\delta P}{1+\bar{S}}  + 2\mathcal{H} \frac{\delta S'}{1+\bar{S}}=0\,.
\end{equation}
Here, $\delta P = c_s^2\, \delta \rho$ and $\delta S'$ can be obtained from differentiating Eq.~\eqref{delta_s_sol} with respect to conformal time. Combining Eqs.~\eqref{delta_s_sol}--\eqref{R1} in this way, we obtain one main equation for the temporal evolution of scalar field $\phi$ with conformal time $\eta$. 

On the other hand, with $\psi = \phi$ and $\mathcal{H}\, \delta S = \bar{S}' \phi$, Eq.~\eqref{possion_raw} can be expressed as
\begin{equation}
(\mathcal{H} + \ss) \,\Delta \phi\, +\Delta \phi' = \frac{3}{2}(1+w)\,\mathcal{H}^2\, \theta\,.
\end{equation}
Substituting for $\phi'$ the formula given in Eq.~\eqref{R1}, we find
\begin{equation}
2 \Delta^2 \,\phi +3 \mathcal{H}\,\ss \,\Delta \phi = 9(1+w)\,\mathcal{H}^3\, \theta - \frac{\kappa\,a^2}{1+\bar{S}}\Delta(\delta\rho)\,, 
\end{equation}
or, in terms of spatial Fourier transforms as defined in Eq.~\eqref{Q2}, 
\begin{equation}
\tilde{\phi}(\eta, \mathbf{k}) = \frac{9(1+w)\,\mathcal{H}^3\, \tilde\theta(\eta, \mathbf{k}) + \kappa\,a^2\,k^2\,\widetilde{\delta\rho}(\eta, \mathbf{k})/(1+\bar{S})}{2k^4 - 3  \mathcal{H}\,\ss k^2}\,.
\end{equation}

%!!!!!!!!!!!!!!!!!!!!!!!!!!!!!!!!!!!!!!!!!!	
\subsubsection{Vector Sector}
The vector field $\mathbf{v}$ is now the only nonzero vectorial degree of freedom in our model. It satisfies
\begin{equation}\label{R2}
\Delta v_i - 6(1+w)\,\mathcal{H}^2\,(v_i+\mathfrak{u}_i^{\perp}) = 0\,
\end{equation}
and 
\begin{equation}\label{R3}
v'_i=-\left(2\mathcal{H}+\ss\right)\,v_i\,.
\end{equation}
As $2\mathcal{H}+\bar{S}'/(1+\bar{S})$ is always a positive function of conformal time, the vectorial degree of freedom has a decaying evolution and eventually becomes negligible as the universe expands. 

The components of the spatial velocity perturbation can be obtained from Eqs.~\eqref{0iraw1} and~\eqref{R2}, namely, 
\begin{equation}
\frac{3}{2} (1+ w) \mathcal{H}^2\, \mathfrak{u}_i^{\parallel} = \partial_i [ (\mathcal{H} + \ss)\phi + \phi']\,, 
\end{equation}
\begin{equation}
6 (1+ w) \mathcal{H}^2 \,\mathfrak{u}_i^{\perp} = \Delta v_i - 6(1+w)\,\mathcal{H}^2\,v_i\,.
\end{equation}
Hence, $\mathfrak{u}_i^{\perp}$ decays in conformal time with the expansion of the universe. 

%!!!!!!!!!!!!!!!!!!!!!!!!!!!!!!!!!!!!!!!!!!
\subsubsection{Tensor Sector}
The tensor field  $\chi_{ij}$ occurs only in the perturbation Eq.~\eqref{ijraw-sym}. The tensor part decouples from the other perturbation equations and we find $\Theta_{ij}=0$. That is, 
\begin{equation}\label{R4}
\chi_{ij}^{''} + \left(2\mathcal{H}+\ss\right)\chi_{ij}^{'}-\Delta\,\chi_{ij} = 0\,.
\end{equation}
A similar equation for free gravitational waves in Minkowski spacetime was studied in~\cite{Tabatabaei:2022tbq} within the framework of the local limit of NLG. In the present work, we are interested in the $H_0$ tension and ignore the possibility of existence of gravitational waves in our model; therefore, we assume  $\chi_{ij} = 0$ in the rest of this paper. 

We have worked out the basic field equations governing the perturbations of the tetrads and the energy-momentum tensor in order to determine the spectrum of CMB temperature anisotropy. In the next subsection, we further develop the dynamical equations governing $\delta$ and $\theta$.

%@@@@@@@@@@@
%@@@@@@@@@@@
\subsection{Continuity and Euler Equations}

From the solutions of the field Eqs.~\eqref{M11} and~\eqref{M12} for the background modified Cartesian flat model, we find
\begin{equation}\label{PP}
	\bar{\rho}'=-3\,(1+w)\,\bar{\rho}\,\mathcal{H} - \frac{3\mathcal{H}^2}{\kappa \, a^2} \,\bar{S}'\,.
\end{equation} 
This gives us the simple solution $\bar{\rho} 	\propto a^{-3(1+w)}/(1+\bar{S})$~\cite{Tabatabaei:2022tbq}. Moving on to the perturbative part of the energy-momentum content of our model,  we note that it is possible to find two conservation formulas for perturbed quantities in the energy-momentum tensor as well.  In principle the conservation equations are not independent of field equations obtained in previous subsections. Indeed, the situation here is rather similar to GR \cite{Amendola:2015ksp}. 

Let us define as in Eq.~\eqref{T5},
\begin{equation}
\mathcal{E}^{\mu}{}_{\hat{\alpha}} := \sqrt{-g}\,(T_{\hat{\alpha}}{}^\mu + \mathcal{T}_{\hat{\alpha}}{}^\mu -\frac{\Lambda}{\kappa}\,e^\mu{}_{\hat{\alpha}})\,;
\end{equation}
then, the continuity equation follows from   ${\partial}_\mu \mathcal{E}^{\mu}{}_{\hat{0}} = 0$, while  $\partial_\mu \mathcal{E}^{\mu}{}_{\hat{i}} = 0$ results in the Euler equation in this case. Using these results or the perturbation equations in the scalar sector, we find the following equations for the density contrast  $\tilde{\delta}$ and volume expansion $\tilde{\theta}$ both in Fourier space:
\begin{equation}\label{convcon1}
	\tilde{\delta}' + A_1\,\tilde{\delta} + A_2 \,\tilde{\theta} = 0\,, 
\end{equation}
\begin{equation}\label{convcon2}
	\tilde{\theta}' + B_1\,\tilde{\theta} + B_2 \,\tilde{\delta} = 0\,. 
\end{equation}
Here, the functions $A_1$, $A_2$, $B_1$ and $B_2$ are defined as
\begin{small}
	\begin{equation}
		A_1 = -\frac{18\,(1+w)\mathcal{H}^3
			+4k^2 \ss + 3\mathcal{H}^2\left(7-3w-6\,\delta P/\delta \rho\right) \ss +12\,k^2 \mathcal{H} (w - \delta P/\delta\rho) -  6 \mathcal{H}\ss'}{2(1+\bar{S})\mathfrak{K}}\,,
	\end{equation}
\end{small}
\begin{small}
	\begin{equation}
		A_2 = -a\frac{-4k^4+54(1+w)\mathcal{H}^4+12 \mathcal{H}\,\ss(2\,k^2 + 3\mathcal{H}^2)+9\,\mathcal{H}^2\left[2k^2(1+w)-\ss^2-2\ss'\right]
		}{2\,k^2\,(1+\bar{S})\mathfrak{K} (1+w)^{-1}}\,,
	\end{equation}
\end{small}
\begin{small}
	\begin{equation}
		B_1 =-(2+3\,w)\mathcal{H}-\ss+\frac{3\mathcal{H}[2\,k^2+3(1+w)\mathcal{H}^2]
		}{\mathfrak{K} }\,,
	\end{equation}
\end{small}
\begin{small}
	\begin{equation}
		B_2 =k^2\frac{ -2k^2\,\,\delta P/\delta\rho +3(1+w)\mathcal{H}^2+3(1+\delta P/\delta\rho)\,\ss\,\mathcal{H}
		}{(1+w)\mathfrak{K} }\,,
	\end{equation}
\end{small}
where
\begin{equation}\label{GK}
\mathfrak{K}  :=2k^2 - 3\,\ss\,\mathcal{H}\,.
\end{equation} 

Solving  Eqs.~\eqref{convcon1} and~\eqref{convcon2}, we have the density contrast for each redshift and wavenumber; accordingly, the matter power spectrum is known. To study the dynamics of perturbations in this theory and find the angular power spectrum of temperature anisotropy of CMB,  four perturbed quantities $\phi$, $\psi$, $\delta$ and $\theta$ are in general needed. These can be obtained by solving equations \eqref{phipsirel}, \eqref{R1}, \eqref{convcon1} and \eqref{convcon2}.  The \texttt{CAMB} can solve these equations in the synchronous gauge. Therefore, in the next subsection, we write the above equations in the synchronous gauge and Fourier space.

 %@@@@@@@@@@@
%@@@@@@@@@@@
\subsection{Cosmological Perturbations in the Synchronous Gauge} 

To confront our model with cosmological observations, the perturbed gravitational field equations must be numerically integrated to determine the temporal evolution of the model with the expansion of the universe. To compare the result with observational data, it is convenient to use the synchronous gauge. This gauge has been explored in \cite{Ma:1994dv}. Therefore, we need to transform the perturbation equations from the conformal Newtonian gauge to the synchronous gauge in order to be able to use the \texttt{CAMB} code~\cite{Lewis:1999bs}. This code makes it possible to find the CMB angular power spectrum. 

In the synchronous gauge,  the metric takes the form
\begin{equation}
 ds^2 = a^2(\eta)\left[-d\eta^2 + (\delta_{ij} + \texttt{h}_{ij})dx^i\,dx^j  \right]\,.
\end{equation}
Here,  $\texttt{h}_{ij}$ has the Fourier decomposition 
	\begin{equation}\label{PSG1}
		\texttt{h}_{ij}(\eta, \mathbf{x}) = \int{d^3 k \, e^{i\,\mathbf{k}.\mathbf{x}}\left[\hat{k}_i \, \hat{k}_j\, \tilde{h}(\eta, \mathbf{k}) + 6\,(\hat{k}_i \, \hat{k}_j - \frac{1}{3}\, \delta_{ij})\, \tilde{\mathfrak{h}}(\eta, \mathbf{k})\right]}\,,
	\end{equation} 
where $\tilde{h}$ and $\tilde{\mathfrak{h}}$ are scalar functions. 
	The coordinate transformation that can take the conformal Newtonian gauge to synchronous gauge is of the general form
	\begin{equation}\label{PSG2}
		x^\mu \rightarrow \hat{x}^\mu = x^\mu + b^\mu (x^\nu)\,.
	\end{equation} 
It can be shown that, using the proper vector field $b^\mu$, the   perturbed quantities $\tilde{\psi}$ and $\tilde{\phi}$ in  the conformal Newtonian gauge can be transformed to the synchronous gauge.  We note  that the quantities with tilde are in Fourier space. The results are:
		\begin{equation}\label{PSG3}
		\tilde{\psi}{}^{\rm{(syn)}}(\eta,\mathbf{k}) = \frac{1}{2\,k^2}\left\{\tilde{h}^{''}(\eta,\mathbf{k}) + 6\, \tilde{\mathfrak{h}}^{''}(\eta,\mathbf{k}) + \frac{a'}{a}\left[\tilde{h}'(\eta,\mathbf{k}) + 6\,\tilde{\mathfrak{h}}(\eta,\mathbf{k})\right]\right\}\,,
	\end{equation} 
	\begin{equation}\label{PSG4}
		\tilde{\phi}{}^{\rm{(syn)}}(\eta,\mathbf{k}) = \tilde{\mathfrak{h}}(\eta,\mathbf{k}) - \frac{1}{2\,k^2}\,\frac{a'}{a}\left[\tilde{h}'(\eta,\mathbf{k})+6\,\tilde{\mathfrak{h}}'(\eta,\mathbf{k})\right]\,.
	\end{equation} 

	We  should also transform the energy-momentum tensor components as
	\begin{equation}\label{PSG6}
		T^\mu{}_\nu\ {}^{\rm{(syn)}} = \frac{\partial \hat{x}^\mu}{\partial x^\sigma}\,\frac{\partial  x^\rho}{\partial\hat{x}^\nu} T^\sigma{}_\rho {}^{\rm{(con)}},
	\end{equation}
	where ``syn" represents the synchronous gauge and ``con" the conformal Newtonian gauge, which will appear as superscripts on perturbed quantities as below: 
	\begin{equation}\label{PSG7}
		\tilde{\delta}^{\rm{(syn)}} = \tilde{\delta}^{\rm{(con)}}-\Upsilon\frac{\bar{\rho}'}{\bar{\rho}}\,,
	\end{equation}
	\begin{equation}\label{PSG8}
		\widetilde{\delta P}^{\rm{(syn)}} = \widetilde{\delta P}^{\rm{(con)}}-\Upsilon \bar{P}'\,,
	\end{equation}
	\begin{equation}\label{PSG9}
		\tilde{\theta}^{\rm{(syn)}} =  \tilde{\theta}^{\rm{(con)}}-\Upsilon k^2\,.
	\end{equation}
	Here, $\Upsilon = \tfrac{1}{2\,k^2}\,(\tilde{h}'+6\tilde{\mathfrak{h}}')$.
Then, the resulting modified perturbation equations become
	\begin{equation}\label{PSG10}
		k^2 \tilde{\mathfrak{h}} - \frac{1}{2}\mathcal{H} \left(\tilde{h}'-3\,\ss\,\tilde{\mathfrak{h}}\right) = -\frac{1}{2}\kappa\,a^2 \frac{\widetilde{\delta\rho}^{(syn)}}{1+\bar{S}}\,,
	\end{equation}
	\begin{equation}\label{PSG11}
		k^2\left(\tilde{\mathfrak{h}}' + \ss\, \tilde{\mathfrak{h}} \right) = -\frac{3}{2}(1+w)\tilde{\theta}^{(syn)}\mathcal{H}^2\,,
	\end{equation}
	\begin{equation}\label{PSG12}
		\tilde{h}^{''}+2\mathcal{H}\,\tilde{h}'-\left(2k^2 + 3\,\ss\,\mathcal{H}- 6\,\ss^2 +\,6\,\ss' \right)\tilde{\mathfrak{h}}
		= \frac{3\kappa\,a^2}{1+\bar{S}}\,\widetilde{\delta P}^{(syn)}+\frac{18}{k^2}(1+w)\tilde{\theta}^{(syn)}\,\ss\,\mathcal{H}^2 \,,
	\end{equation}
	\begin{equation}\label{PSG13}
		2k^2\,\tilde{\mathfrak{h}} = 2\mathcal{H}\left(\tilde{h}'+6\,\tilde{\mathfrak{h}}'\right)+\tilde{h}^{''}+6\,\tilde{\mathfrak{h}}^{''}\,.
	\end{equation}
	
	Let us now  express the density contrast  $\tilde{\delta}$ and volume expansion $\tilde{\theta}$ in this gauge, namely, 
	\begin{small}
	\begin{multline}\label{PSG14}
		\tilde{\delta}'^{(syn)} = -(1+w)\left(1-\frac{9}{2k^2}\mathcal{H}\ss\right)\tilde{\theta}^{(syn)} - \frac{(1+w)k^2+\mathcal{H} \ss+\ss^2 - \ss'}{\mathfrak{L}}\:\tilde{h}'\\
		+\frac{\ss\left[2k^2-\frac{3}{2}\left(1-3w+6\frac{\delta P}{\delta\rho}\right)\mathcal{H}^2\right]+3\mathcal{H}\left[2k^2\left(w-\frac{\delta P}{\delta\rho}\right)+2\ss ^2-\ss'\right]
		}{\mathfrak{L}}\,\tilde{\delta}^{(syn)},
	\end{multline}
\end{small}	
where $\mathfrak{L}  :=2k^2+3\,\ss\,\mathcal{H}$ and
		\begin{equation}\label{PSG15}
			\tilde{\theta}^{(syn)\prime} = -\left((1-3w)\mathcal{H} - \ss\right)\tilde{\theta}^{(syn)}+\frac{k^2}{1+w}\frac{\delta P}{\delta \rho}\, \tilde{\delta}^{(syn)}+\frac{1}{\mathcal{H}}\,\ss\,\frac{k^2}{1+w} \tilde{\mathfrak{h}}\:.
		\end{equation}
Finally we need to relate the parameter $\delta P$ to $\delta \rho$. This is usually done with defining the sound speed, $c_s^2 := (\delta P/\delta \rho) |_{\rm{Fluid Rest Frame}}$. In the synchronous gauge, however,  the perturbed equation of state  becomes
\begin{equation}\label{dp}
\widetilde{\delta P}=c_{s}^2\widetilde{\delta\rho}+3\mathcal{H}\bar{\rho}(1+w)(c_{s}^2-w)\frac{\tilde{\theta}}{k^2}\,.
\end{equation}
In this way, the required perturbation equations are in the synchronous gauge and we can obtain the CMB angular power spectrum using the \texttt{CAMB} code.

%%**************************************%%
\begin{figure} 
	\includegraphics[scale=0.65]{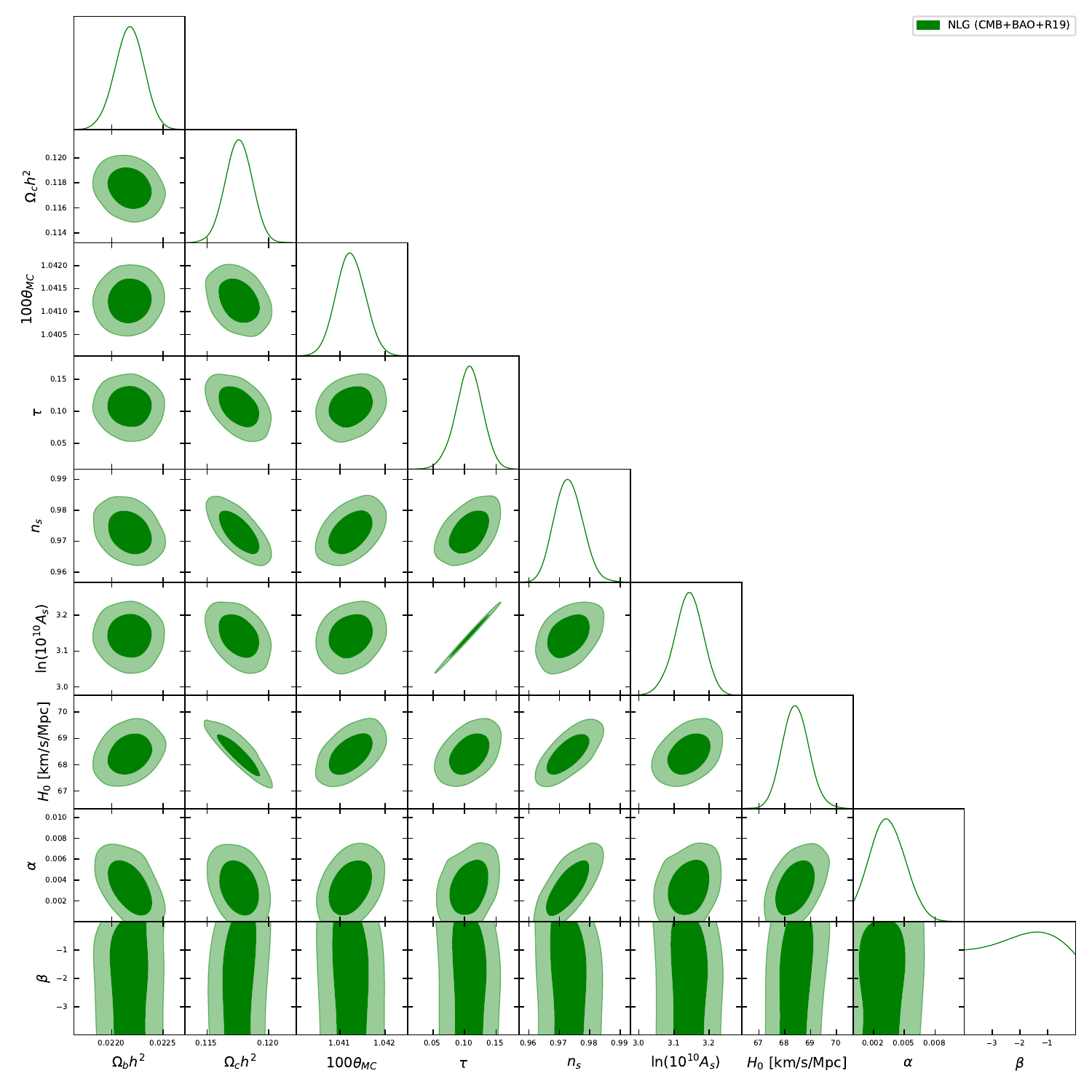} 
	\caption{{Two-dimensional posterior probabilities for the free parameters of the modified Cartesian flat model with the susceptibility function $S$ described by Eq.~\eqref{bg}. The model has been constructed within the framework of the local limit of nonlocal gravity (NLG). These constraints  are obtained by using CMB, BAO and R19.  This plot is produced by employing the publicly available code \texttt{GetDist} \cite{Lewis:2019xzd}. The dark smaller regions show the $1\sigma$ and the lighter broader regions indicate the $2\sigma$ confidence levels.}} \label{fig1}
\end{figure} 
%%**************************************%%
%@@@@@@@@@@@@@@@@@@@@
%@@@@@@@@@@@@@@@@@@@@
%@@@@@@@@@@@@@@@@@@@@
%@@@@@@@@@@@@@@@@@@@@
\section{Results}
\label{sec5}

The modified Cartesian flat cosmological model under consideration here is a solution of the local limit of nonlocal gravity theory. We have worked out the perturbations of this model in the synchronous gauge. Accordingly, by choosing a specific susceptibility function $S(t)$, we can compare the theory with the observational data. As proposed and investigated in~\cite{Tabatabaei:2022tbq}, we choose 
\begin{equation}\label{bg}
S(z) = \alpha \left(1+z\right)^\beta\,,
\end{equation}
where $z$ is the cosmological redshift and $\alpha$ and $\beta$ are two extra free parameters in addition to the ones in common with the $\Lambda$CDM model.
Hence, the set of free parameters of our modified model is as follows:
\begin{equation}
\mathcal{P}=\{\Omega_b\,h^2,\Omega_c\,h^2,100\,\Theta_{MC},\tau,\ln[10^{10}A_s], n_s,\alpha,\beta\}\,,
\end{equation}
where $\Omega_bh^2$ and $\Omega_ch^2$ are the current physical baryonic and dark matter densities respectively; $\Theta_{MC}$ is the acoustic angular scale on the CMB sky; $\tau$ is the re-ionization optical depth, $A_s$ is the amplitude of primordial power spectrum of scalar perturbations, inherited from inflation; $n_s$ is the spectral index of those primordial perturbations and describes their scale dependence;
 $\alpha$ shows the deviation of our model away from $\Lambda$CDM and finally $\beta$ indicates the  dependence of $S$ on the cosmological redshift.
In order to find the best-fitting values for the free parameters, we need to use observational data and define a so-called loss function $\chi^2$, which quantifies the deviation of our model's prediction from the data. Naturally, the goal is to minimize $\chi^2$, which is a function of the free parameters of the model. Best-fitting values for the free parameters are those which minimize $\chi^2$. Usually, the loss function in cosmology is not an analytic function of the free parameters and one cannot perform differentiation to find the minimum. This is because the predictions of a model for an observable like CMB temperature angular power spectrum consist of many numerical integrations and there is no simple analytic form of the prediction as a function of free parameters. In these cases, in order to have the minimum of the loss function, one has to check every point in the discretized space of the free parameters, $\mathcal{P}$, and find the point where the loss function is a minimum. This procedure is resource intensive, especially when the dimension of $\mathcal{P}$ (number of free parameters) is more than $\sim 5$. On the other hand, the relevant region in $\mathcal{P}$, where the loss function is close to its minimum, is a rather small zone and can be found quickly with a few steps in a kind of \textit{biased} random walk in $\mathcal{P}$. In such a random walk, every step is biased toward the direction in which $\chi^2$ decreases. After taking a sufficient number of such steps in $\mathcal{P}$, one can find the best point; moreover, one has now access to a chain of points on which every step had landed during the random walk.  These chains can be used to show the confidence level of the result. The statistically 1-2 $\sigma$ levels of confidence are related to the points at which steps landed during the walk within $68\%$ and $95\%$ confidence intervals, respectively. Accordingly,  the density of these points at each region gives the relevance of that region. More dense zones are more probable ones for the free parameters. The resulting probability distribution for the free parameters is called the posterior probability. Often, one projects the posterior probability onto one and two dimensional subspaces of $\mathcal{P}$, in order to have a sensible view of it  and to infer the constraints on each parameter.  
In this regard, we made use of ($i$) the CMB temperature and polarization angular power spectra, reported by the Planck team~\cite{Planck:2015bpv}, referred to as ``CMB"; ($ii$) the measurement of $H_0$ by Riess et al.~\cite{Riess:2019cxk}, referred to as ``R19"; ($iii$) distances to several baryon acoustic oscillations, serving as standard rulers located at redshift $\lesssim 1$~\cite{Beutler:2011hx,Ross:2014qpa,Blake:2011en}, referred to as ``BAO".  We put both background and perturbation equations into the \texttt{CAMB} code \cite{camb,Lewis:1999bs} to have the observational implications of this model and using the \texttt{CosmoMC} code \cite{Lewis:2002ah,Lewis:2013hha}, we sampled the space of free parameters and obtained the desired chain of points. In Figure \ref{fig1} and Table \ref{tab}, the resulting posterior probabilities for these parameters are shown. It is evident there is a positive correlation between $\alpha$ and $H_0$. So in this model, CMB is more compatible with the local measurements of the $H_0$. It is worth noticing that $\alpha$ is more than 1$\sigma$ away  from zero, where our modified model reduces to $\Lambda$CDM.  It also seems that the  data employed here is insensitive to the value of $\beta$ and this quantity has not been constrained, which shows that while the data prefers $S$ to increase from $S=0$ at $z=\infty$ to $S=\alpha$ at the present epoch, the shape of this transition is not significant.

\begin{table}
		\begin{tabular}{ c c c }
			\hline&&\\
			$\quad$Parameter$\quad$&$\quad$Prior\,Values$\quad$&$\quad$Posterior\,Values$\quad$\\
			\hline&&\\
			$\Omega_b\,h^2$&[0.005\ ,\ 0.1]&$0.02217\pm 0.00014$\\&&\\
			
			$\Omega_c\,h^2$&[0.001\ ,\ 0.99]&$0.1176\pm 0.0011$\\&&\\
			
			100\,$\Theta_{MC}$&[0.5\ ,\ 10]&$1.04124\pm 0.00032$\\&&\\
			
			$\tau$&[0.01\ ,\ 0.8]&$0.107\pm 0.021$\\&&\\
			
			$n_s$&[0.8\ ,\ 1.2]&$0.9730^{+0.0043}_{-0.0049}$\\&&\\
			
			$\ln[10^{10}A_s]$&[2\ ,\ 4]&$3.142\pm 0.041$\\&&\\
			
			$\alpha$&[0\ ,\ 0.5]&$0.0034^{+0.0016}_{-0.0019}$\\&&\\

			$\beta$&[-4\ ,\ 0]&$---$\\&&\\
			
			$H_0\ [\rm km/s/Mpc]$&-&$68.42\pm 0.52$\\\hline&&\\\hline
		\end{tabular}		
	\caption{\label{tab:best-fit}The best-fitting values and $68\%$ confidence level intervals for the parameters of the modified model when CMB, BAO and R19 are used. The Prior Values column, shows the interval in which every parameter is explored.} \label{tab}
\end{table}

In Table \ref{tab}, we summarize the best-fitting parameters of our model with the prior values we used and the 1$\sigma$ confidence levels. 

To have a closer look at the behavior of $S(z)$ at low redshifts, it proves useful to adopt a reconstruction approach. In fact, reconstructing various cosmic functions based on Gaussian processes can give substantial constraints on the possible models emerging from teleparallel gravity; see, for instance, \cite{Briffa:2020qli}.
	The main rationale behind reconstructing a cosmic curve is to infer a \textit{model-independent} function for a specific observable from a discrete set of data points. For instance, consider the Hubble parameter at a few redshifts provided by cosmological surveys.  One may be willing to find the most \textit{reasonable} continuous curve that goes through them. This goal can be achieved by binning the redshift space and letting the $H_{\rm Rec}(z)$ at each bin to be a Gaussian random  variable with a mean value which best fits the data. However, to avoid over-fitting and to ensure smoothness, it is necessary to demand a sort of correlation between $H_{\rm Rec}(z)$'s at different redshifts encoded in a kernel. The assumed correlation should depend on $|z_1-z_2|$ and also  have the property of fading at redshift separations larger than a certain correlation length $l$. There are plenty of kernels satisfying these properties, such as $\mathcal{C}_1(z_1,z_2)=\mathcal{C}(0)\exp\big[- (z_1-z_2)^2/(2\,l^2)\big]$ or $\mathcal{C}_2(z_1,z_2)=\mathcal{C}(0)/\big[1+(z_1-z_2)^2/l^2\big]$ and so on, where $\mathcal{C}(0)$ sets the strength of correlation.       
	
\begin{figure}[h]
	\centering
	\includegraphics[width=\linewidth]{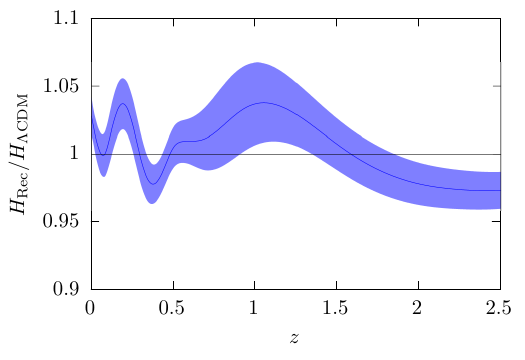}
	\caption{Reconstructed $H(z)$ associated with 1$\sigma$ uncertainty, normalized to that of the $\Lambda$CDM model, done in~\cite{Wang:2018fng}. In  their reconstruction procedure~\cite{Wang:2018fng}, the following data sets have been used: CMB distance information from Planck \cite{Planck:2015fie}, the ``Joint Light-curve Analysis" (JLA) supernovae \cite{SDSS:2014iwm}, BAO measurements from 6dF Galaxy Survey (6dFGS) \cite{Beutler:2011hx}, SDSS DR7 Main Galaxy Sample (MGS)\cite{Ross:2014qpa},  tomographic BOSS DR12 (TomoBAO) \cite{BOSS:2016zkm}, eBOSS DR14 quasar sample (DR14Q) \citep{Ata:2017dya} and the Lyman-$\alpha$ forest (Ly$\alpha$FBAO) of BOSS DR11 quasars \cite{BOSS:2013igd, BOSS:2014hwf}, the local measurement of $H_0=73.24 \pm 1.74 \rm \,[km\,s^{-1}\,Mpc^{-1}]$ in \cite{Riess:2016jrr}, and the Observational Hubble parameter Data (OHD) \cite{Moresco:2016mzx}.
	}
	\label{fig:recpdfh}
\end{figure}

\begin{figure}
	\centering
	\includegraphics[width=\linewidth]{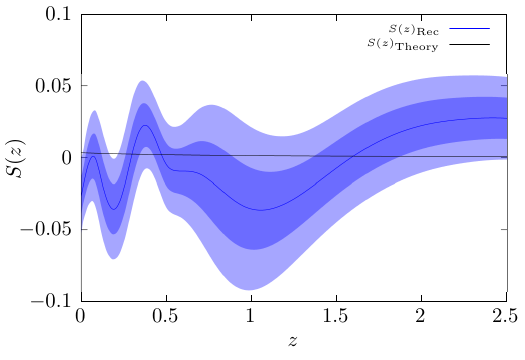}
	\caption{Inferred $S(z)_{\rm Rec}$ with its 1$\sigma$ and 2$\sigma$  confidence regions.   $S(z)_{\rm Theory}$ is the nearly horizontal curve in the middle of the figure and is given by the function $S(z)= \alpha (1+z)^\beta$ with $\alpha=0.005$ and $\beta=-1.3$.  }
	\label{fig:recpdfsnn}
\end{figure}

	In this work, we employ the previously reconstructed $H(z)$ by Wang et al. in~\cite{Wang:2018fng} as depicted in Figure \ref{fig:recpdfh}. The data used in the reconstruction method is mentioned in the caption of Figure \ref{fig:recpdfh}. Wang et al.~\cite{Wang:2018fng} have followed the method of Crittenden et al.~\cite{Critt, Crittenden:2011aa}, which employed Gaussian processes with a correlation of the form $\mathcal{C}(a_1,a_2)=\mathcal{C}(0)/\big[1+(a_1-a_2)^2/a^2_c\big]$ between $H_{\rm Rec}(a)$'s at two different scale factors. They have also adopted $a_c=0.06$ for the correlation length and $\mathcal{C}(0)=0.0085$ for the correlation strength. It is important to emphasize the model-independent nature of reconstruction and the fact that no model or theory can affect this process because it is basically a non-parametric approach in cosmology and hence is unaffected by theoretical assumptions. However, it can affect models by putting constraints on them. In our case, we have the phenomenological behavior of $H(z)$ up to $z\sim2.5$ and the functional form of the reconstructed  $S(z)$ can be derived via Eq.~\eqref{A14} of  Appendix A, namely, 
	\begin{equation}\label{GP1}
		S = \mathfrak{R}^{1/2} \frac{H_{\Lambda \rm{CDM}}}{H} - 1\,, 
	\end{equation}
	where $\mathfrak{R}$ is given by Eq.~\eqref{A11}. That is, 
	\begin{equation}\label{GP2}
		\mathfrak{R} = \frac{\rho_i(t_0)}{\rho_{i}^{\Lambda \rm{CDM}}(t_0)}\,(1+\alpha)\,,
	\end{equation}
	where we have used the fact that $S(z = 0) = \alpha$. In our analysis, the parameters for   $S(z)_{\rm Theory}$ are $\alpha=0.005$ and $\beta=-1.3$. Assuming that energy densities of various components of the two models agree at the present epoch and the fact that $(1+\alpha)^{1/2} \approx 1.0025$, we can ignore the deviation of $\mathfrak{R}$ from unity and use the reconstructed form for the $H(z)$ to infer the reconstructed $S(z)$ from
	\begin{eqnarray}
		S(z)_{\rm Rec}=\frac{H_{\rm \Lambda CDM}}{H_{\rm Rec}}-1\,,
	\end{eqnarray}
	and its error, induced by the uncertainty in the reconstructed $H(z)$, as
	\begin{eqnarray}
		\Delta S(z)_{\rm Rec}=\big[1+S(z)_{\rm Rec}\big]\frac{\Delta H_{\rm Rec}}{H_{\rm Rec}}\,.
	\end{eqnarray}
	The shape of $S(z)_{\rm Rec}$ alongside with the best fit of the assumed model described with Eq.~\eqref{bg} is shown in  Figure \ref{fig:recpdfsnn}. It seems that $S(z)_{\rm Rec}$ behaves in a more complicated manner than what our power-law model is capable of describing. However, it oscillates around $S(z)_{\rm Theory}$ in such a way that $S(z)_{\rm Theory}$ (solid black line in the middle of Figure \ref{fig:recpdfsnn}) almost everywhere lies within the 2$\sigma$ confidence region of the $S(z)_{\rm Rec}$. This means that our simplified model of $S(z)$ is compatible with the model-independent reconstruction of $S(z)$.

\section{Discussion}
\label{sec6}

In this paper, linear perturbations about the modified Cartesian flat cosmological solution of the local limit of nonlocal gravity theory are derived and investigated. The modified model differs from $\Lambda$CDM through a susceptibility function $S(t)$.  By choosing this function as in Eq.~\eqref{bg}, a new cosmological model is specified. Using CMB temperature and polarization angular power spectra, the viability of this model has been verified. Our results show that this model fits the data well and is to some extent capable of reconciling CMB data with the local measurements of the Hubble constant.

\section*{Acknowledgments}
	
We would like to thank Levon Pogosian for providing us with the reconstructed $H(z)$~\cite{Wang:2018fng}.
The work of A. B.  has been financially supported by the Iran Science Elites Federation. The work of S. B. is partially supported by the Abdus Salam International Center for Theoretical Physics (ICTP) under the regular associateship scheme. Moreover, S. B. and J. T. are partially supported by the Sharif University of Technology Office of Vice President for Research under Grant No. G4010204.

\appendix

\section{Connection between the modified and $\Lambda$CDM models}

The purpose of this appendix is to establish a connection between the modified Cartesian flat model and the standard $\Lambda$CDM cosmological model. Equations~\eqref{M11}--\eqref{M12} for the unperturbed modified Cartesian flat model can be expressed in terms of cosmic time $t$ as
\begin{equation}\label{A1}
3 (1+S) \left(\frac{\dot a}{a}\right)^2 = 8 \pi G  [\rho + \Lambda/(8 \pi G)]\,, 
\end{equation} 
\begin{equation}\label{A2}
2(1+S) \frac{\ddot a}{a} + (1+S) \left(\frac{\dot a}{a}\right)^2 = - 8 \pi G [P-\Lambda/(8\pi G)] - 2 \frac{dS}{dt}\frac{\dot a}{a}\,. 
\end{equation} 

When $S = 0$, we recover the standard equations of the $\Lambda$CDM model, namely, 
\begin{equation}\label{A3}
3 \left(\frac{\dot a}{a}\right)^2 = 8 \pi G \sum_i \rho_{i}^{\Lambda \rm{CDM}}(t)\,, 
\end{equation} 
\begin{equation}\label{A4}
2 \frac{\ddot a}{a} +  \left(\frac{\dot a}{a}\right)^2 = - 8 \pi G \sum_i P_{i}^{\Lambda \rm{CDM}}(t)\,, 
\end{equation} 
where $P_{i}^{\Lambda \rm{CDM}} = w_i \,\rho_{i}^{\Lambda \rm{CDM}}$ and the perfect fluid components consist of matter (dark matter and baryons), radiation and dark energy. It is possible to reduce the dynamics of the $\Lambda$CDM model to the equation of motion~\eqref{A3}, where 
\begin{equation}\label{A5}
	\rho_{i}^{\Lambda \rm{CDM}}(t)=\rho_{i}^{\Lambda \rm{CDM}}(t_0)\,a(t)^{-3(1+w_i)}\,.
\end{equation}

Let us now return to the modified model. As explained in detail in Tabatabaei et al. (2022), the existence of $S(t)$ with $dS/dt \ne 0$ is incompatible with the presence of a cosmological constant. Thus we set $\Lambda = 0$ in Eqs.~\eqref{A1}--\eqref{A2}; instead, the modified model contains a dynamic dark energy component. In this case, it is possible to choose $w_{de} = -1$ and~\citep{Tabatabaei:2022tbq}
\begin{equation}\label{A6}
	\rho_{de}(t)=\rho_{de}(t_0)\frac{1+S(t_0)}{1+S(t)}\,.
\end{equation}
Moreover, in a similar way as in the $\Lambda$CDM model, we can reduce the dynamics of the modified model to 
\begin{equation}\label{A7}
3 (1+S) \left(\frac{\dot a}{a}\right)^2 = 8 \pi G \sum_i \rho_{i}(t)\,, 
\end{equation}
where
\begin{equation}\label{A8}
	\rho_i(t)=\rho_i(t_0)\frac{1+S(t_0)}{1+S(t)}a(t)^{-3(1+w_i)}\,.
\end{equation}
It follows from Eqs.~\eqref{A5} and~\eqref{A8} that 
\begin{equation}\label{A9}
\frac{	\rho_i(t)}{\rho_{i}^{\Lambda \rm{CDM}}(t)}=\frac{\rho_i(t_0)}{\rho_{i}^{\Lambda \rm{CDM}}(t_0)}\,\frac{1+S(t_0)}{1+S(t)}\,.
\end{equation}
Hence, we can write 
\begin{equation}\label{A10}
	\rho_i(t)= \rho_{i}^{\Lambda \rm{CDM}}(t)\,\frac{\mathfrak{R}}{1+S(t)}\,,
\end{equation}
where $\mathfrak{R}$ is a constant given by 
\begin{equation}\label{A11}
\mathfrak{R} = \frac{\rho_i(t_0)}{\rho_{i}^{\Lambda \rm{CDM}}(t_0)}\,[1+S(t_0)]\,.
\end{equation}
From equations of motion~\eqref{A3} and~\eqref{A7} in the two models, we note that the corresponding Hubble parameters are connected by
\begin{equation}\label{A12}
[1+S(t)] \left(\frac{H}{H^{\Lambda \rm{CDM}}}\right)^2 = \frac{\sum_i \rho_{i}(t)}{\sum_i \rho_{i}^{\Lambda \rm{CDM}}(t)}\,,
\end{equation}
where Eq.~\eqref{A10} implies 
\begin{equation}\label{A13}
 \frac{\sum_i \rho_{i}(t)}{\sum_i \rho_{i}^{\Lambda \rm{CDM}}(t)} = \frac{\mathfrak{R}}{1+S(t)}\,.
\end{equation}
Therefore, from Eqs.~\eqref{A12} and~\eqref{A13}, we obtain the final result, namely,  
\begin{equation}\label{A14}
1+S (t) = \mathfrak{R}^{1/2} \frac{H^{\Lambda \rm{CDM}}(t)}{H(t)}\,.
\end{equation}

%@@@@@@@@@@@@@@@@@@@@@@@@@
%@@@@@@@@@@@@@@@@@@@@@@@@@
%@@@@@@@@@@@@@@@@@@@@@@@@@			

\end{document}